\documentclass{format/aa}

\usepackage[T1]{fontenc}
\usepackage{txfonts}
\usepackage{amsmath,graphicx,booktabs}
\usepackage{hyperref}
\hypersetup{
  colorlinks   = true,
  urlcolor     =[RGB]{255, 104, 68},
  linkcolor    =[RGB]{255, 104, 68},
  filecolor    =[RGB]{255, 104, 68},
  citecolor    =[RGB]{255, 104, 68}
}

\usepackage{xspace}
\newcommand{\snr}{\ensuremath{\mathrm{S/N}}\xspace}
\defcitealias{chen2022}{C22}
\makeatletter\newcommand{\figurecaptionhere}{\def\@captype{figure}}\makeatother

\begin{document}

\title{Machine-learned candidates for hidden repeaters among CHIME/FRB one-off bursts are often grating-lobe detections}
\titlerunning{Hidden repeaters and grating lobes}

\author{Dany~Vohl\inst{1}\fnmsep\thanks{E-mail: vohl@astron.nl}}
\institute{ASTRON, Netherlands Institute for Radio Astronomy, Oude Hoogeveensedijk 4, 7991 PD Dwingeloo, The Netherlands}
\authorrunning{D.~Vohl}

\date{Received ---; accepted ---}

\abstract
{Many papers apply machine learning to CHIME/FRB catalogue data to find apparently one-off fast radio bursts that are repeaters in hiding, and several list candidates.
Their most important features are spectral, because repeater bursts are generally narrowband.}
{We test whether these lists depend on where a burst fell in CHIME's beam, where a broadband burst detected in a grating lobe of the formed beams loses part of its band and looks narrowband.}
{We flag grating-lobe detections among the $127$ one-off bursts with baseband data by comparing two positions for each burst: the catalogue position, which takes the localization solution nearest the beam with the strongest detection, and the more accurate baseband position.
We compare the flags with eleven published candidate lists and the second CHIME/FRB catalogue.}
{Of the $127$ one-off bursts with baseband positions, we identify $37$ ($29\%$) detected in a grating lobe, consistent with the $(30 \pm 4)\%$ reported by CHIME/FRB from the same data.
For each, the offset between the two positions predicts the frequency at which the burst's spectrum peaks.
The same holds for a fifth to a quarter of the bursts of known repeaters.
One classification lists $89\%$ of these bursts as candidates, against $16\%$ of the main-lobe bursts, and its two-dimensional embedding places them next to known repeaters. 
Grating-lobe detections are over-represented in all eleven lists (nine papers, partly overlapping).
The association between candidates and other one-off bursts does not come from brightness as they have similar signal-to-noise ratios.
In the second catalogue, that classification's candidates repeat more often than other one-off bursts once CHIME's exposure is allowed for, through sources confirmed by 2023, but no list anticipates later repeaters better than chance, and too few bursts repeat to compare the lobes.}
{Among the brighter bursts with baseband data, much of each list can be explained by position in the beam, which is unrelated to repetition.
Survey classifications should be tested against beam position before being interpreted.}


\maketitle

\section{Introduction}
\label{sec:intro}

Only a minority of fast radio bursts have been seen to repeat, and whether the apparently one-off bursts form a separate population, or are repeaters whose repeat bursts were missed or are to be detected, bears directly on their origin.
Of the sources detected by CHIME/FRB up to September 2023, $2.4\%$ have been seen to repeat. 
The limits on how often the one-off sources repeat lie within the range of the repeaters' rates, and therefore the rates do not provide evidence for two populations \citep{cook2026}.
Repeater bursts have been shown to differ in morphology: they are wider in time and narrower in frequency than one-off bursts \citep{pleunis2021}.
Since the first CHIME/FRB catalogue \citep{chimecat1}, many papers have used that difference with machine learning, supervised or unsupervised, to look for repeaters among the one-off bursts \citep[e.g.][\citealt{kharel2026}]{chen2022, luo2023, zhuge2023, sharma2024, garcia2024frb, sun2025, qiang2025, madheshwaran2025, junior2026, mankatwit2026}.
Several publish the one-off bursts that their classification places with the known repeaters.
The features these classifications rank as most important are mostly spectral---spectral running, bandwidth, peak or highest frequency---and \citet{luo2023} rank brightness temperature and rest-frame bandwidth as the two most important.

A spectrum measured by CHIME is not only a property of the burst.
CHIME forms four beams east--west from its four cylinders \citep{chimefrbsystem}, whose $22$\,m spacing \citep{chimeoverview} gives the beams grating lobes that move with frequency. 
A broadband burst detected in a grating lobe loses part of its band and appears to be narrowband (Appendix~\ref{app:model}).
\citet{pleunis2021} noted that bursts detected away from the centre of a formed beam lose part of their band, but could not correct for it because the first-catalogue positions are too imprecise, and argued that it does not explain the difference between repeaters and one-off bursts because both are subject to nearly identical biases.
Bursts of known repeaters are indeed displaced by a lobe almost as often as one-off bursts (Section~\ref{sec:checks}), but a bias shared by both classes can still move individual one-off bursts towards the repeaters, and that is what a classification picks up.
\citet{chimecat2} attribute most of the steep spectral running among apparent non-repeaters to this effect, and quote, from analysis of baseband data for $129$ catalogue bursts, that about $30\%$ were detected in a grating lobe, without listing which bursts they are.
Here we identify them from public data and set them against the published candidate lists.

Section~\ref{sec:data} describes the data.
Section~\ref{sec:results} shows that the candidates follow position in the beam, rules out alternative explanations, and verifies if they later repeated.
Section~\ref{sec:discussion} discusses implications of the finding for classifications of survey data, before concluding in Section~\ref{sec:conclusions}.

\section{Data}
\label{sec:data}

We use the first CHIME/FRB catalogue from VizieR (J/ApJS/257/59), one row per burst (its first detection burst), with the spectral running\footnote{Spectral running is the curvature term of the running power law fitted to each burst's spectrum, and is strongly negative for narrowband bursts.}, peak frequency, bandwidth and signal-to-noise ratio (\snr) of the catalogue's burst fits.
Precise positions for $140$ catalogue bursts come from the baseband update \citep{chimebaseband}.
Repeaters confirmed after the first catalogue come from \citet{chimerep2023}, and the assignment of bursts to repeating sources up to September 2023 comes from the table of the second catalogue \citep{chimecat2}, which we obtain through CANFAR (doi:10.11570/25.0066).
The candidate lists are those published by \citet[hereafter C22]{chen2022}, \citet{zhuge2023}, \citet{luo2023} (two lists, from their combined classifier and from a nearest-neighbour classifier), \citet{sharma2024}, \citet{sun2025}, \citet{qiang2025}, \citet{madheshwaran2025}, \citet{junior2026} (two lists, from their primary features and from primary and derived features) and \citet{mankatwit2026}.
The full list of \citet{sharma2024} is obtained from their supplementary material, and $59$ of its $66$ bursts are in the first catalogue; \citet{madheshwaran2025} classify only bursts with baseband data.
The labels of \citetalias{chen2022} cover $468$ of the $474$ one-off bursts; we count a burst as their candidate if any of its rows carries that label, which gives $182$ candidates (noting their paper quotes $188$ candidates counted differently).
We use the classification of \citetalias{chen2022} as the primary test case because it is the earliest of the lists, made before the 2023 repeater confirmations, and includes a label for every burst.
We do however evaluate all eleven lists in the same manner.

\section{Results}
\label{sec:results}

\subsection{Candidates follow the beam}
\label{sec:beam}

\begin{figure*}
\centering
\includegraphics[width=\textwidth]{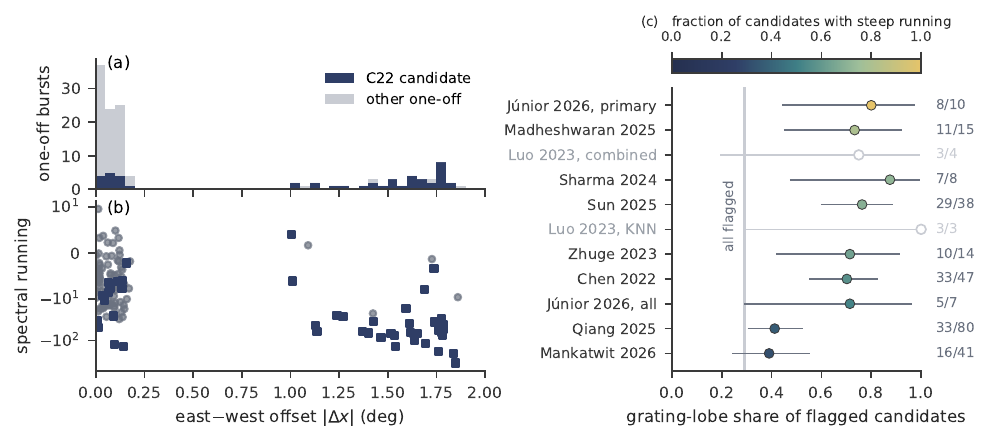}
\caption{One-off CHIME/FRB bursts with baseband positions.
\textbf{(a)} East--west offset between the baseband and catalogue positions: bursts detected in the main lobe sit near zero, and those detected in a grating lobe sit one lobe spacing away.
Dark bars are bursts listed as repeater candidates by \citetalias{chen2022}.
\textbf{(b)} Spectral running against the same offset (steeper running is more negative); 
squares are the candidates of \citetalias{chen2022} and circles the other one-off bursts.
\textbf{(c)} For each list of Table~\ref{tab:frb}, the share of its flagged candidates that were detected in a grating lobe, with the $95\%$ interval and the counts at right; the vertical line, $0.29$, is the share expected for a list with no preference between lobes, equal to the grating-lobe fraction of all $127$ one-off bursts with baseband positions.
Rows are ordered, and coloured, by the fraction of each list's candidates with steep spectral running (below $-35$); the two lists with fewer than five flagged candidates are shown in grey.}
\label{fig:frb}
\end{figure*}

For one-off bursts, the catalogue positions come from intensity data and are the localization solution nearest the beam that detected the burst most strongly; that solution lies in the beam's main lobe even when the burst arrived through a grating lobe \citep{chimecat1}.
Transformed to the telescope frame at the time of arrival, the catalogue positions cluster on CHIME's four columns of formed beams (median distance from the nearest column $0.03\degr$, against $0.10\degr$ for positions spread evenly), between $-0.87$ and $+0.48\degr$ east--west.
The baseband positions resolve the lobe.
For the $127$ one-off bursts with baseband positions, the east--west offset between the catalogue and baseband positions, $\Delta x = \Delta\alpha\cos\delta$, is cleanly bimodal (Figure~\ref{fig:frb}a): at most $0.18\degr$ for $90$ bursts, and $1.0\degr$ to $1.9\degr$, one lobe spacing, for $37$.
Computing the offset in the telescope frame instead, which differs from $\Delta x$ by up to $0.35\degr$ for bursts near the pole, moves no burst from one group to the other.
One further one-off burst lies $7.4\degr$ away, in a far sidelobe, and is excluded.
Known repeaters are excluded because the first catalogue lists them at their source positions rather than per burst; Section~\ref{sec:checks} tests them with the per-burst positions of the second catalogue.
The resulting grating-lobe fraction, $29\%$, agrees with the $(30 \pm 4)\%$ that \citet{chimecat2} quote from CHIME's analysis of the same baseband data.

The offsets match the expectation from grating lobes.
A first grating lobe of CHIME's four cylinders, spaced $d = 22$\,m apart \citep{chimeoverview}, lies at $\sin\theta = c/(f d)$ from the main lobe, so the frequency at which a burst's offset places the lobe should track the frequency at which the burst was detected.
For the $37$ grating-lobe bursts, the larger the offset, the lower the catalogue peak frequency (Spearman $\rho = -0.76$).
The frequency predicted from each burst's offset differs from its peak frequency by a median of $29$\,MHz.
Pairing the same predicted and peak frequencies at random gives a median of $79$\,MHz ($5$ to $95\%$ range $54$ to $108$\,MHz), and none of $10^4$ random pairings comes as close.
Instead of a single frequency, the composite beam model of CHIME/FRB \citep{chimecat1} predicts the spectrum that a flat-spectrum burst at the baseband position would show in the detecting beam, which we take to be the beam with the highest response at the catalogue position.
The peak of the predicted spectrum lies a median of $21$\,MHz from the catalogue peak frequency for the $37$ grating-lobe bursts, against $77$\,MHz for the $90$ main-lobe bursts. 
Hence, the beam model predicts the spectrum only for bursts that came through a grating lobe, and an error in localisation would not produce this relation.
Of the $37$, $34$ arrived at east--west positions outside the span of every main-lobe burst.

Among the $127$ one-off bursts with a lobe flag, $68\%$ of the $37$ grating-lobe detections have steep spectral running (below $-35$, the division used by \citealt{chimecat2}), against $3\%$ of the $90$ main-lobe detections, and their median bandwidths are $181$ and $400$\,MHz.
Of the $28$ flagged one-off bursts with steep running, $25$ are grating-lobe detections (Figure~\ref{fig:frb}b).
The two groups do not differ in \snr~(medians $31.4$ and $31.3$).

Table~\ref{tab:frb} sets the lobe flag against the eleven candidate lists.
\citetalias{chen2022} call $33$ of the $37$ grating-lobe detections candidates, and $14$ of the $90$ main-lobe detections (Fisher's exact test, $p = 4\times10^{-15}$).
In every list, grating-lobe detections are over-represented among the candidates; after correcting for eleven lists, the excess is significant in seven.
Grating-lobe detections make up $39$ to $100\%$ of each list's flagged candidates, against $29\%$ of all $127$ one-off bursts with baseband positions, the share expected for a list with no preference between lobes.
The two lists with the lowest shares, those of \citet{qiang2025} and \citet{mankatwit2026}, are also the two with the fewest steep-running candidates ($37$ and $32\%$, against $51$ to $100\%$ for the others), as expected if the lobe reaches the classifications through the spectrum (Figure~\ref{fig:frb}c).
The eleven lists are not independent confirmations: they partly share features and candidates, and two papers give two lists each.
By the ratio of intersection to union, the candidates of \citetalias{chen2022} and \citet{sun2025} overlap by $0.65$, and no other two lists overlap by more than $0.46$.
As the methods share the same catalogue features, the excess in each list is not independent of the others.

\begin{table*}
\centering
\caption{Published repeater candidates among one-off CHIME/FRB bursts whose beam lobe is known from baseband positions ($37$ in a grating lobe, $90$ in the main lobe).
The fourth column is the share of each list's flagged candidates that were detected in a grating lobe, with its $95\%$ interval; a list with no preference between lobes would have $0.29$, the grating-lobe fraction of all $127$ one-off bursts with baseband positions.
The last column is the probability from Fisher's exact test, multiplied by eleven to account for testing eleven lists.}
\label{tab:frb}
\begin{tabular}{lrrlr}
\toprule
Candidate repeaters list & Grating & Main & Share & $p \times 11$ \\
\midrule
\citetalias{chen2022}               & 33/37 & 14/90 & 0.70 [0.55, 0.83] & $4\times10^{-14}$ \\
\citet{zhuge2023}              & 10/37 & 4/90  & 0.71 [0.42, 0.92] & $0.007$ \\
\citet{luo2023}, combined      & 3/37  & 1/90  & 0.75 [0.19, 0.99] & $0.81$ \\
\citet{luo2023}, KNN           & 3/37  & 0/90  & 1.00 [0.29, 1.00] & $0.26$ \\
\citet{sharma2024}             & 7/37  & 1/90  & 0.88 [0.47, 1.00] & $0.008$ \\
\citet{sun2025}                & 29/37 & 9/90  & 0.76 [0.60, 0.89] & $9\times10^{-13}$ \\
\citet{qiang2025}              & 33/37 & 47/90 & 0.41 [0.30, 0.53] & $9\times10^{-4}$ \\
\citet{madheshwaran2025}       & 11/37 & 4/90  & 0.73 [0.45, 0.92] & $0.002$ \\
\citet{junior2026}, primary    & 8/37  & 2/90  & 0.80 [0.44, 0.97] & $0.009$ \\
\citet{junior2026}, all        & 5/37  & 2/90  & 0.71 [0.29, 0.96] & $0.24$ \\
\citet{mankatwit2026}          & 16/37 & 25/90 & 0.39 [0.24, 0.55] & $1$ \\
\bottomrule
\end{tabular}
\end{table*}

\subsection{Checks}
\label{sec:checks}

We test the robustness of the result for \citetalias{chen2022} against three alternative explanations in which associations would arise from brightness, from the steepest spectra alone, or from uncertain lobe flags.
The association does not come from brightness given that in the fainter half of the flagged sample, $16$ of $19$ grating-lobe and $4$ of $45$ main-lobe bursts are candidates ($p = 7\times10^{-9}$), and in the brighter half $17$ of $18$ and $10$ of $45$ ($p = 10^{-7}$); nor are the candidates fainter overall.
Among the one-off bursts, the candidates of \citetalias{chen2022} and the other one-off bursts have similar distributions of fitted \snr, fluence and peak flux; the probability that a random candidate exceeds a random non-candidate is $0.45$ to $0.55$ for all three parameters, and no difference is significant ($p \geq 0.06$).
The same holds for the \snr at detection ($0.49$).
Candidates and other one-off bursts differ strongly in spectral running and bandwidth, where the same probability is $0.17$ and $0.21$.
The candidate fraction is $30\%$ in the faintest quarter of the one-off bursts and $40$ to $44\%$ in the others, the opposite of what noise would produce, since noise narrows the measured band of faint bursts.
It also does not come only through the steepest spectra. 
Among bursts with spectral running above $-35$, $8$ of $12$ grating-lobe and $11$ of $87$ main-lobe bursts are candidates ($p = 10^{-4}$).
\citet{chimebaseband} mark two of the flagged one-off bursts as having an uncertain lobe, and three as detected more than $2\degr$ from the meridian, where the primary beam is not modelled; removing all five gives $31$ of $33$ against $14$ of $89$.

We check whether bursts of known repeaters, which are intrinsically narrowband, are also placed one lobe spacing away.
The first catalogue lists each repeater at its source position, but the second gives a position for every burst, from the real-time data, again taking the solution nearest the beam that detected it most strongly \citep{chimecat2}.
We compare each repeat burst with the position of its source, the baseband position for $13$ sources and the median of the burst positions for $46$ others with at least three bursts.
Of $929$ repeat bursts, from $59$ sources, $257$ lie one lobe spacing away east--west, from $45$ sources, and $657$ lie within $0.5\degr$.
For the lobe-displaced bursts, the frequency at which the offset places the grating lobe differs from the catalogue peak frequency by a median of $20$\,MHz, against $54$\,MHz when peak frequencies are shuffled among bursts of the same source and $78$\,MHz when they are shuffled across sources; moving the offsets of main-lobe bursts out by one lobe spacing gives no agreement.
The share of lobe-displaced bursts falls with \snr, from $31\%$ below \snr~$\leq10$ to $14\%$ above $30$; weighted to the \snr of the one-off bursts with baseband data gives $22\%$, against $29\%$ for one-off bursts, a difference that is not significant ($p = 0.13$ above \snr~$15$).
In $33$ of the $42$ sources with bursts in both lobes, the lobe-displaced bursts have the narrower median bandwidth.
These data cannot tell whether the lobe narrows the bursts or whether narrowband bursts are the ones detected through a lobe.

All shares in Table~\ref{tab:frb} refer to the bursts with baseband data, which are brighter than the rest of the catalogue (median \snr $31$ against $19$); their fractions of steep spectral running ($22\%$) and of candidates ($37\%$) are slightly lower than in the rest at comparable brightness ($28$ and $43\%$).

\subsection{Grating-lobe bursts location in the UMAP embedding}
\label{sec:embedding}

\citetalias{chen2022} present the coordinates of every burst in the two-dimensional embedding in which they found their clusters, which allows us to measure distances inside their classification.
In that embedding, grating-lobe one-off bursts lie a median $0.18$ units from the nearest known-repeater burst, against $3.08$ for main-lobe one-off bursts (Mann--Whitney $p = 6\times10^{-12}$).
Across all one-off bursts, the distance to the nearest repeater follows spectral running (Spearman $\rho = 0.68$), and among bursts with shallow running the grating-lobe bursts still sit closer to the repeaters ($0.84$ against $3.11$, $p = 4\times10^{-4}$).

Without knowing their exact clustering settings and random seed, our attempt to reproduce their embedding from the catalogue agrees poorly with their labels. 
Therefore, results above use their published coordinates.
In $29$ of our $30$ reproduction runs, grating-lobe bursts were candidates more often than main-lobe bursts; without the spectral features, they were in only $5$ of $10$.

\subsection{Did the candidates repeat?}
\label{sec:hits}

Six one-off FRBs of the first catalogue were later confirmed as the first bursts of repeaters \citep[in the gold sample of][]{chimerep2023}.
Three predictive lists were published before that confirmation, namely those of \citetalias{chen2022}, \citet{zhuge2023} and \citet{luo2023}. 
\citetalias{chen2022} included five of these six as candidate repeaters, which with a list of its size would occur by chance with probability $0.03$; \citet{zhuge2023} listed one ($p = 0.54$); and \citet{luo2023} listed none.
Correcting for the three lists yields $p \approx 0.1$.
Of the five bursts listed by \citetalias{chen2022}, FRB~20190110C was itself detected in a grating lobe, FRB~20190430C and FRB~20190609C in the main lobe, and the remaining two do not have a published baseband position.
\citetalias{chen2022} list none of the six silver one-off bursts of \citet{chimerep2023}, where a random list of the same size would be expected to include two or three ($p = 0.05$).

From the second CHIME/FRB catalogue \citep{chimecat2}, bursts are assigned to repeating sources where applicable.
The catalogue covers the period up to September 2023.
Among them, $11$ of the one-off bursts of the first catalogue are assigned to a repeater.
Six are the gold one-off bursts of \citet{chimerep2023}, and a seventh, FRB~20190201A, is a later burst of the gold source FRB~20181226F.
The other four belong to repeaters first reported by \citet{cook2026}: FRB~20190208C, FRB~20190224D and FRB~20190320E in their gold sample, and FRB~20181018B in their silver sample.
A twelfth, FRB~20190329A, is assigned to the known repeater FRB~20180814A, and hence we exclude it; it is one of the $182$ candidates of \citetalias{chen2022}. 
This leaves $181$ candidates against $292$ other one-off bursts, including the six unlabeled by \citetalias{chen2022}.
The catalogue assigns a burst to a repeater only when the grouping of bursts is unlikely to be a chance coincidence, so a burst not yet assigned may still repeat.

Of the $11$ one-off bursts that later repeated, $7$ are among the $181$ candidates of \citetalias{chen2022}, with the remaining $4$ among their $292$ bursts labelled as one-off events (one-sided Fisher $p = 0.08$).
Fewer than one in $25$ of their candidates had repeated by 2023 September.
This comparison understates the success of their list.
Given CHIME's longer exposure to sources near the pole, repeat bursts are more often detected there.
All four non-candidates that repeated lie at declinations of $72\degr$ to $89\degr$, where non-candidates outnumber candidates at about a two to one ratio.

We verify two avenues of allowing for declination by permuting the labels of \citetalias{chen2022} within declination bands, and by drawing the repeaters in proportion to each burst's exposure in the first catalogue.
Both avenues agree, with the permutation giving $p = 0.05$, and the draw giving $p = 0.04$.
The excess lies below declination $70\degr$, where $5/139$ candidates of \citetalias{chen2022} later repeated, all of them gold sources, against $0/211$ bursts they did not list.
Above this declination, $2/42$ candidates repeated, against $4/81$ other bursts not in their list.
The other two forecast lists show no excess.
The later lists of \citet{qiang2025}, \citet{madheshwaran2025} and one of \citet{junior2026} do (declination-adjusted $p = 0.003$ to $0.04$), but only through the gold sources already confirmed as repeaters when these lists were published.

Given all eleven lists were published prior to the second catalogue, we can test every list with the first five one-off bursts later assigned to a repeater.
We find that none predicted better than chance: \citetalias{chen2022} listed $2$ of the $5$, against $1.9$ expected, and the best list, that of \citet{qiang2025}, $4$, against $2.9$ expected ($p = 0.29$), although with five bursts only a strong excess would be detected.
Four of the five have shallow spectral running, and the steepest, FRB~20190208C, was detected in a grating lobe.

Too few bursts repeated to tell whether grating-lobe detections repeat less often than main-lobe detections.
Even if every repeat fell in the main lobe, the difference would not be significant.
The second catalogue places all $37$ grating-lobe one-off bursts at least $0.48\degr$ away from their true positions. 
Its grouping of bursts into repeaters nonetheless tolerates east--west offsets of this size \citep{cook2026}, and allowing explicitly for a later burst one lobe spacing away does not add more matches than expected by chance.

\section{Discussion}
\label{sec:discussion}

The excess of grating-lobe detections among the evaluated lists does not show that the candidates are not repeaters. In fact, one list anticipated later repetition, and a grating-lobe detection can well be a repeater.
What the result shows is that, for a large part of each list, the narrow spectrum that made a burst a candidate can be explained by where the burst fell in the beam, and where a burst falls in the beam does not inform on whether a source would repeat or not.

The list of \citetalias{chen2022} anticipated the repeaters confirmed by 2023 (Section~\ref{sec:hits}), but with baseband positions for only three of the five, we cannot tell whether its success comes from main-lobe candidates.
The known repeaters show that the effect is not confined to the candidates.
A fifth to a quarter of their bursts in the second catalogue were detected in a grating lobe, since each burst arrives at a different point in the source's transit, and therefore the repeater class used to train classifiers is itself partly shaped by the beam.
\citet{cook2026} note that detection is biased against the narrow spectra and long durations of repeater bursts.
Grating lobes add a different effect, making broadband one-off bursts look narrowband---a phenomenological aspect of repeater bursts.
Grating-lobe detections resemble repeater bursts in their spectra but not in duration, as \citet{chimecat2} found for one-off bursts with steep spectral running.
In the second catalogue, the $37$ grating-lobe one-off bursts have a median intrinsic width of $0.42$\,ms, against $1.8$\,ms for repeat bursts detected in either lobe (excluding FRB~20220912A and FRB~20201124A).
Detection through a grating lobe therefore does not explain why these bursts are short.
\citetalias{chen2022} include width and scattering time among their $13$ features and list $33$ of these $37$ one-off bursts as candidate repeaters.

The effect is, of course, not peculiar to CHIME.
The primary beam of every radio telescope narrows with increasing frequency, so a burst detected away from the beam centre loses the top of its band and looks narrowband, more weakly than in a grating lobe but in the same direction.
We focus on CHIME here given the published classification works use its catalogue; as other burst samples grow, the same test against beam position also applies.

More generally, a classification of survey data should not only focus observed burst properties but also be tested against every known instrumental property that can shape the measured features like the position in the beam or on the detector, observing configuration, and data-processing version.
Baseband positions have been public since 2023, before seven of the eleven lists, and none of the classifications we examined compared them with the catalogue positions.
\citet{madheshwaran2025} classify the same baseband sample and note that some positions changed from the catalogue, but use the change only to recompute the Galactic dispersion measure.

\section{Conclusions}
\label{sec:conclusions}

We tested published machine-learned candidates for hidden repeaters among CHIME/FRB one-off bursts against the position of each burst in the telescope's beam.
Our main conclusions are as follows:

\begin{enumerate}
\item The offset between each burst's baseband and catalogue positions tells whether it was detected in a grating lobe; this reproduces CHIME's own grating-lobe fraction, and the offset predicts the burst's peak frequency.
\item Grating-lobe detections are over-represented among the candidates of all eleven lists examined, taken from nine papers, making up $39$ to $100\%$ of each list's candidates with baseband data against $29\%$ expected; the excess is significant in seven. 
The candidates do not differ in brightness from other one-off bursts, so this is not an effect of brightness.
\item In the published embedding of one classification, grating-lobe bursts sit next to the known repeaters, closer than main-lobe bursts, including among bursts with shallow spectral running.
\item Grating-lobe detections resemble repeater bursts in their spectra but not in duration, so their short widths are not caused by the lobe.
\item In the repeater assignments of the second CHIME/FRB catalogue, the candidates of one list repeat more often than other one-off bursts once CHIME's exposure is accounted for, through sources confirmed as repeaters by 2023. 
No list anticipates the repeaters first found in that catalogue better than chance, and too few bursts repeat to compare grating-lobe and main-lobe candidates.
\item Because the second CHIME/FRB catalogue tabulates the main-lobe position, as it documents, its positions for grating-lobe detections lie away from the true position---as is the case for all $37$ grating-lobe one-off bursts and a fifth to a quarter of repeat bursts---typically by one lobe spacing, at the offset their peak frequency predicts.
\item Classifications of survey data should be tested against the known instrumental variables before they are interpreted.
\end{enumerate}

\section*{Data availability}
All data used are public: the first CHIME/FRB catalogue (VizieR J/ApJS/257/59), the second CHIME/FRB catalogue (CANFAR, doi:10.11570/25.0066), the baseband positions and repeater tables published by \citet{chimebaseband} and \citet{chimerep2023}, and the candidate lists published by the papers in Table~\ref{tab:frb}.

\begin{acknowledgements}
This research has made use of the VizieR catalogue access tool, CDS, Strasbourg, France \citep{vizier2000}, and of NASA's Astrophysics Data System.
This research used the facilities of the Canadian Astronomy Data Centre operated by the National Research Council of Canada with the support of the Canadian Space Agency.
The analysis used NumPy \citep{numpy2020}, SciPy \citep{scipy2020}, pandas \citep{pandas2010}, Astropy \citep{astropy2022}, scikit-learn \citep{sklearn2011}, UMAP \citep{umap2018} and Matplotlib \citep{matplotlib2007}.
\end{acknowledgements}

\bibliographystyle{format/aa}
\bibliography{refs}

\begin{appendix}
\twocolumn[{%
\section{A model of the grating-lobe effect}
\label{app:model}
\begin{center}
\includegraphics[width=0.93\textwidth]{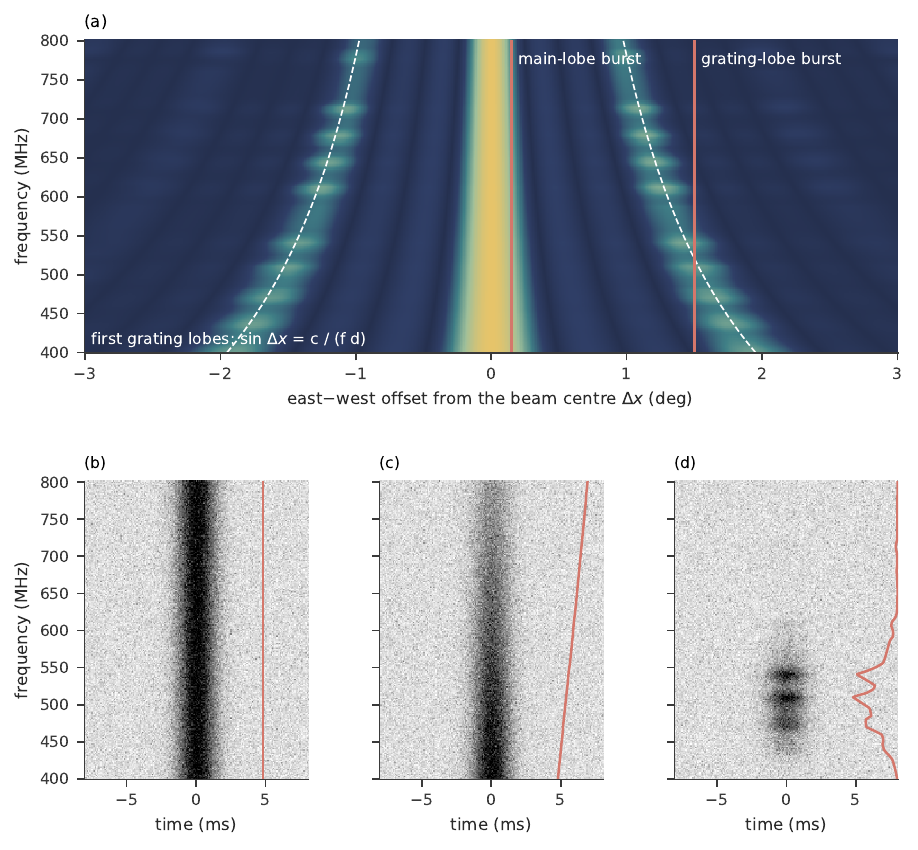}
\end{center}
\figurecaptionhere
\caption{Model of how a grating lobe changes the dynamic spectrum of a broadband burst.
\textbf{(a)} Response of one formed beam in the composite beam model of CHIME/FRB (beam $1128$, near the zenith; \citealt{chimecat1}), relative to the response at the beam centre, against east--west offset from the beam centre and frequency, on a square-root scale; dashed lines mark the first grating lobes, $\sin\Delta x = c / (f d)$, and vertical lines the positions of the two bursts below.
\textbf{(b)} A burst with a flat spectrum from $400$ to $800$\,MHz, with dispersion removed and noise added.
\textbf{(c)} The same burst detected in the main lobe, $0.15\degr$ from the beam centre, rescaled for display.
\textbf{(d)} The same burst detected in the grating lobe, $1.5\degr$ from the beam centre, rescaled for display.
Orange curves are the spectra.}
\label{fig:model}
\vspace{1.5em}
}]

Figure~\ref{fig:model} shows how a grating lobe changes the dynamic spectrum of a broadband burst, for one formed beam of the composite beam model of CHIME/FRB \citep{chimecat1}.
A first grating lobe lies where $\sin\Delta x = c / (f d)$, where $\Delta x$ is the east--west offset from the beam centre, $f$ the frequency, $c$ the speed of light and $d = 22$\,m the spacing of the cylinders, so it moves towards the main lobe as the frequency rises.
A burst $1.5\degr$ east--west of the beam centre therefore falls in the grating lobe only near $510$\,MHz, where the beam passes a slice about $85$\,MHz wide at half maximum, with up to $38\%$ of the gain at the beam centre.
In the beam model the lobe is broken into knots along frequency, which make the spectrum uneven within the slice (Figure~\ref{fig:model}d).
The burst then looks narrowband, and its catalogue position, which takes the main-lobe solution, is $1.5\degr$ from the true one.
The first grating lobe crosses the CHIME band from $800$\,MHz at an offset of about $0.95\degr$ to $400$\,MHz at about $1.9\degr$, so a broadband burst detected in a grating lobe can look narrowband anywhere in the band.
A burst detected this way keeps $3.5$ to $12\%$ of the band-integrated signal it would have at the beam centre, so it must be intrinsically about $8$ to $30$ times brighter to be detected through a grating lobe than at the beam centre.
\end{appendix}

\end{document}